\documentstyle[12pt,clustering,psfig]{article}
\def\mg{\big <}
\def\md{\big >}
\def\d{{\rm d}}

\def\be{\begin{equation}}

\def\ee{\end{equation}}

\def\hkpc{$h^{-1}$kpc }
\def\hMpc{$h^{-1}$Mpc }
\def\h3Mpc{h^{-3}{\rm Mpc}^3 }

\def\h3Mpcinv{h^{3}{\rm Mpc}^{-3} }

\def\spose#1{\hbox to 0pt{#1\hss}}
\def\simlt{\mathrel{\spose{\lower 3pt\hbox{$\mathchar"218$}}
     \raise 2.0pt\hbox{$\mathchar"13C$}}}
\def\simgt{\mathrel{\spose{\lower 3pt\hbox{$\mathchar"218$}}
     \raise 2.0pt\hbox{$\mathchar"13E$}}}

\begin{document}
\heading{RECENT ADVANCES IN GRAVITATIONAL LENSING}

\centerline{Y. Mellier$^{(1,4)}$, L. Van Waerbeke$^{(2)}$, F. Bernardeau$^{(3)}$,
B. Fort$^{(4)}$}
{\it
\centerline{$^{(1)}$Institut d'Astrophysique de Paris,}
\centerline{98$^{bis}$ Boulevard Arago,}
\centerline{75014 Paris, France.}
\centerline{$^{(2)}$ Observatoire Midi Pyr\'en\'ees,}
\centerline{14 Av. Edouard Belin,}
\centerline{31400 Toulouse, France.}
\centerline{$^{(3)}$ Service Physique Th\'eorique,}
\centerline{CE de Saclay,}
\centerline{91191 Gif-sur-Yvette  Cedex, France.}
\centerline{$^{(4)}$Observatoire de Paris (DEMIRM),}
\centerline{61 Av. de l'Observatoire,}
\centerline{75014 Paris, France.}
}

\begin{abstract}{\baselineskip 0.4cm 
The gravitational lensing effect is one of the most promising tools for 
cosmology. Indeed it probes directly the total
mass distribution in large-scale structures and can
as well provide valuable informations on 
the values of the density parameter $\Omega$ and 
the cosmological constant $\lambda$. In this review we summarize the 
most spectacular observational and theoretical advances obtained 
in this field during the last five years. 
}
\end{abstract}

\section{Introduction}

The observed structures of the universe are thought 
to originate from the gravitational 
condensation of primordial mass-energy fluctuations 
whose evolution depends on the power spectrum 
of the initial fluctuations, the amount and  
nature of the various matter components present in the early universe  
and the eventual existence of a cosmological constant. Since 
most of the acting gravitational masses remain invisible, 
constraining cosmological 
scenario from astronomical observations has been one of the most 
fascinating  challenge of the last decades. From 
the observations of the spatial galaxy distribution (see de Lapparent,
this conference), and from dynamical studies of gravitational systems 
like galaxies or clusters of galaxies, it is possible to infer the
amount of dark matter as well as its distribution, or
to have insights into the clustering properties of 
the visible matter in the universe. 
However, none of these observations measures {\sl directly} the total
amount and the distribution of matter: angular or redshift surveys give the
distribution of {\sl light} associated with galaxies, and mass estimations of
gravitational structures can only be obtained within the assumption that
they are simple
relaxed or virialised dynamical systems. Dynamics of galaxy velocity fields
seems an efficient and promising approach to map directly the potentials
responsible from large-scale flows, but catalogs are still poor and some
assumptions are still uncertain or poorly understood on a physical
basis. 

Gravitational lensing effects are  fortunately a direct probe of
deflecting masses in the universe. They allow to determine 
directly the amount of matter present along the line-of-sight
from observed or reconstructed deflection angles.  After the 
discovery of the first multiply imaged quasar (Walsh et al. 1979) and 
the first observations of gravitational arcs (Soucail et al. 1987, 1988; 
Lynds \& Petrossian 1986) and arclets (Fort et al. 1988), 
gravitational lensing rapidly becomes one of the most useful tool 
for probing dark matter on 
all scales and cosmological parameter as well. In the following, we 
summarize what we have learned from strong and weak lensing regimes 
in clusters 
of galaxies and how constraints
on the cosmological parameters can, or could, be obtained. 
We will also discuss a technique 
for measuring gravitational shear that has been developed
recently and which opens new perspectives for the 
observations and the analysis of lensing effect by large-scale structures, as 
those we discuss in the last section.

\begin{figure}
\vskip -2cm
\hskip 3truecm
\psfig{figure=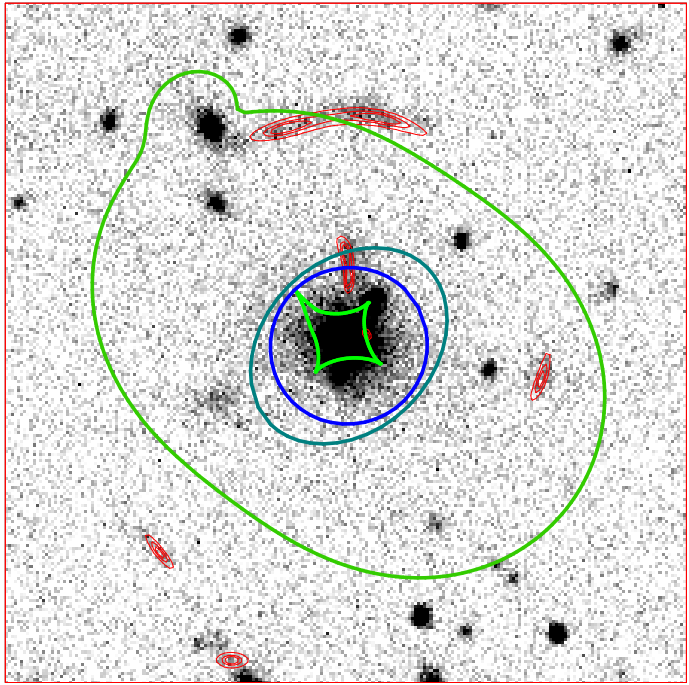,width=10. cm}
\caption{Model of MS2137-23. This cluster is at redshift 0.33 and shows
a tangential arc and the first radial arc ever detected. Up to now it is
the most constrained cluster and the first one where counter images were
predicted before being observed (Mellier et al. 1993). The external and
the dark internal solid lines are the critical lines. The internal grey
ellipse and the diamond are the caustic lines. The thin isocontours
shows the positions of the arcs and their counter images.}
\end{figure}

\section{Definitions and lensing equations}
Let us remind the basic principles of the gravitational lensing effects.
A deflecting mass changes the apparent position of a source  
$\vec \theta_S$ into the apparent image position  $\vec \theta_I$  by 
the quantity $\vec \alpha$:
\be
\vec \theta_S=\vec \theta_I+\vec \alpha(\vec \theta_I)  \ .
\ee
The deflection angle  $\vec \alpha(\vec \theta_I)$ 
is the gradient of the two-dimensional 
(projected along the line of sight at the angular position $\vec \theta_I$) 
gravitational potential $\phi$. 
The gravitational distortion of background objects is
described by the Jacobian of the transformation, namely the amplification
matrix $\cal A$ between the source and the image plane (Schneider, Ehlers \& 
Falco 1992):
\be
{\cal A}=\pmatrix{ 1-\kappa-\gamma_1 & -\gamma_2 \cr
-\gamma_2 & 1-\kappa+\gamma_1 \cr },
\ee
where $\kappa$ is the convergence, $\gamma_1$ and $\gamma_2$ are the
shear components. They are related to the Newtonian gravitational potential
$\phi$ by:
\be
\kappa={1\over 2} \nabla^2 \phi={\Sigma\over \Sigma_{\rm crit.}}; \ \ \  \gamma_1={1\over 2}(\phi_{,11}-\phi_{,22}) \ ; \ \ \ \gamma_2=\phi_{,12} \ ,
\ee
where $\Sigma$ is the projected mass density and $\Sigma_{\rm crit.}$ is the
critical mass density which would exactly focus a light beam originating
from the source on the observer plane.  It depends on the angular diameter 
distances $D_{ab}$, (where $a,b=[o(bserver),l(ens) or s(ource)]$) 
involved in the lens configuration:
\be
\Sigma_{\rm crit.}={c^2\over 4\pi G}{D_{os}\over D_{ol} D_{ls}}  \ .
\ee

Depending on the ratio $l=\Sigma / \Sigma_{\rm crit.}$ 
we most often distinguish
for data analysis the strong 
lensing ($l\gsim1$) and the weak lensing ($l\ll1$) regimes. 
Note that the values of the cosmological parameters enter in the 
relationship between the angular distance and the redshift. This is this 
dependence that, in some cases, can be used to constrain $\Omega$ or
$\lambda$.

\section{Arcs and mass in the central region of clusters of galaxies }

Arcs and arclets correspond to strong lensing cases with $l\gg1$ and 
$l \approx 1$ respectively.  Giant arcs form at the points where the 
determinant of the magnification matrix is (close to) infinite. To these 
points (the critical lines) correspond the caustic lines in the source plane. 
From the position of the source with respect to the caustic line, one can 
easily define typical lensing configuration with formation of giant arcs 
from the merging of two or three images, or radial arcs and 
``straight arc'' as 
well (see Fort \& Mellier 1994, Narayan \& Bartelmann 1996).

\begin{table}
\hskip 0.3 truecm
\psfig{figure=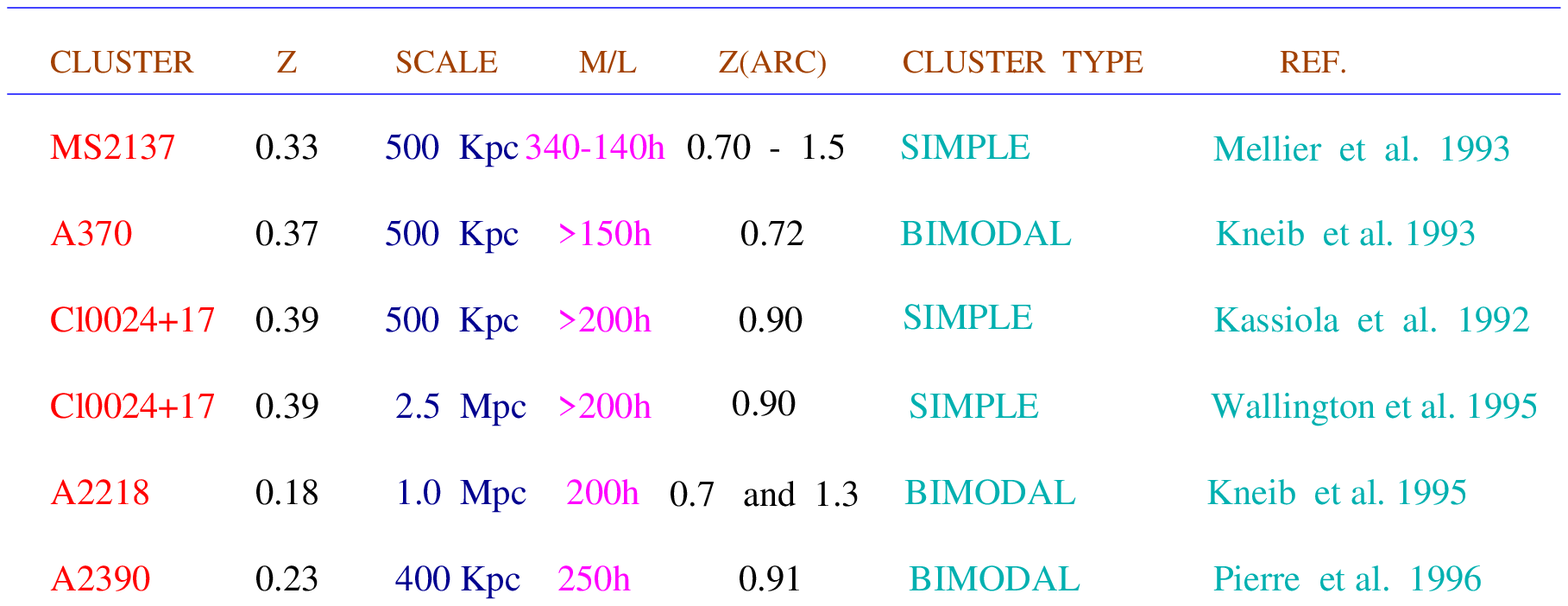,width=16. cm}
\caption{Summary of mass from giant arcs. We only give some typical
examples. More details can be found in Fort \& Mellier (1994) and
Narayan \& Bartelmann (1996). The scales are expessed in $h_{100}^{-1}$
Kpc.}
\end{table}

Some of these typical lensing cases have been observed in rich clusters 
of galaxies. For MS2137-23 (Mellier et al. 1993; see figure 1), A370 
(Kneib et al. 1993), Cl0024+1654 (Wallington et al. 1995), A2218 (Kneib et 
al. 1995), very accurate models with image predictions have been done which 
provide among the most precise informations we have about the 
central condensation of dark matter 
of clusters of galaxies on scale of $\approx 300$\hkpc 
(table 1). In particular, giant arcs definitely demonstrate that 
clusters of galaxies are dominated by dark matter, with 
 mass-to-light ratio (M/L)  larger than 100, which closely follows the 
geometry of the diffuse light distribution associated with the brightest 
cluster members. Furthermore, since arcs 
occur when $\Sigma /\Sigma_{\rm crit.}\gg1$, clusters of galaxies must be much 
more concentrated, e.g. with a smaller core radius, 
than it was expected from their galaxy and X-ray gas 
distributions. Note that the occurrence of arcs is also enhanced by the  
existence of additional clumps observed in most of rich clusters which 
 increases the shear substantially (Bartelmann 1995). The direct 
observations of substructures by using  giant arcs  confirm that clusters 
are dynamically young systems, therefore pointing towards a rather 
high value of $\Omega$.

\section{The weak lensing regime and lensing inversion in clusters}

\subsection{Lensing inversion and mass reconstruction}

Beyond the region of strong lensing, background galaxies are still 
weakly magnified. The light deviation induces a small increase of their 
ellipticity in the direction perpendicular to the gradient of the 
projected potential (shear). Despite the intrinsic ellipticity of the sources 
and some observational or instrumental effects it is possible to 
make statistical analysis of this coherent polarization of the 
images of background galaxies and 
to recover the mass distribution of the lens. 
These weakly distorted galaxies are like a
background distorted grid which can be used to probe the projected mass
density $\Sigma$ of the foreground lens. The  shape parameters 
of the images $M^I$ are related to the shape parameters of the sources
$M^S$ by the equation,

\begin{figure}
\hskip 1truecm
\psfig{figure=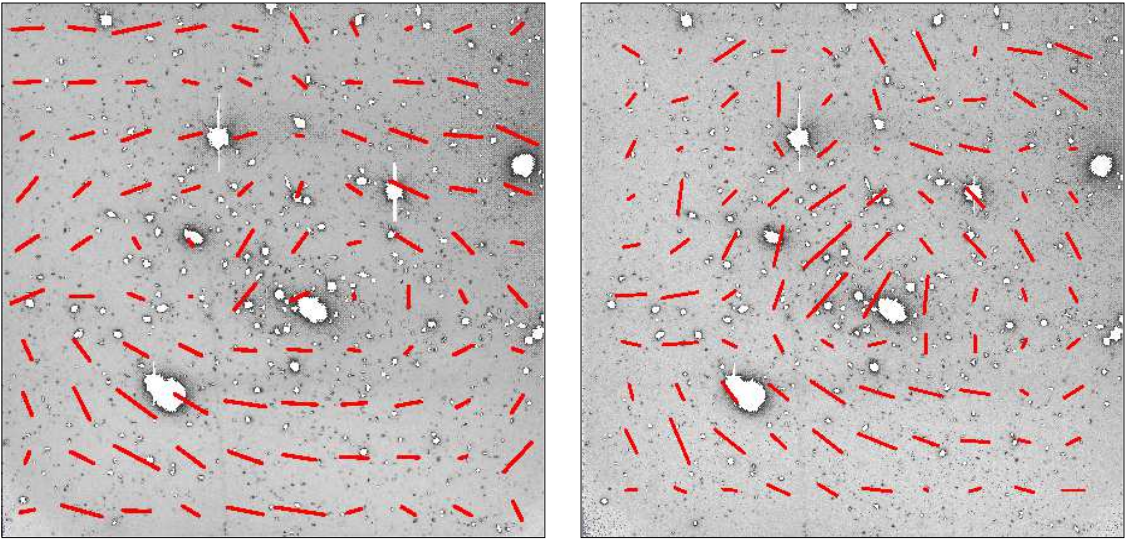,width=15. cm}
\caption{Shear maps around the rich lensing cluster A1942. The CCD images
is a 4 hours exposure obtained at the Canada-France-Hawaii Telescope 
in excellent seeing conditions (0.65"). North is top and east is on the
left. On the left panel, 
the shear is measured by using
the Bonnet \& Mellier's method which consists is computing shape
parameters of an annular aperture centered of each individual galaxies. 
On the right panel, the shear is obtained by computing the
autocorrelation (ACF) of the images (see section 4.2). Though the shape is
almost the same, the signal to noise is higher with the ACF.
Note the
central pattern which shows the bimodal nature of the mass distribution
located on the two brightest cluster members, and the eastern extension
of the shear pattern probably due to a substructure.}
\end{figure}

\be
M^{S}= {\cal A} M^I {\cal A},
\ee
where $M$ are the second order momenta of objects and ${\cal A}$ 
the magnification matrix. We thus have a 
relation between the ellipticity of the sources 
$\vec \epsilon_S$ and the observed ellipticity of the images 
$\vec \epsilon_I$: 
\be
\vec \epsilon_S=
{\vec \epsilon_I-\vec g \over 1-\vec g \cdot \vec \epsilon_I} \ ; 
\ \vec g={\vec \gamma\over 1-\kappa}\ .
\ee
In particular, if the sources are randomly distributed then their 
averaged intrinsic ellipticity verifies  
$\mg\vec \epsilon_S\md=0$ and we have
\be
\vec g=\mg\vec \epsilon_I\md\ .
\ee
Since the main domain of application of the weak lensing is
the large-scale distribution of the Dark Matter, at scale above
$>0.5$ \hMpc, the analysis has been mainly focussed 
on the weak lensing regime, where $(\kappa,
\gamma) \ll 1$. The relations between the physical ($\vec\gamma$) and
observable ($\vec\epsilon_I$) quantities are then simpler,
since we have,
\be
\mg\vec\epsilon_I\md= \vec\gamma\ .
\ee
The projected mass density $\Sigma$ of the lens can be obtained from the
distortion field by using Eq.(8) and the integration of Eq. (3)
(Kaiser \& Squires 1993):
\be
\kappa(\vec \theta_I)={-2\over \pi} \int \d^2{\vec \theta}\ {\vec\chi(\vec
\theta-\vec \theta_I) \over (\vec \theta-\vec
\theta_I)^2} \cdot\vec\gamma(\vec \theta_I) +\kappa_0 \ ; \ \ \ \vec \chi(\vec \theta)=({\theta_1^2-
\theta_2^2\over \theta^2} \ , \  {2\theta_1 \theta_2 \over \theta^2}),
\ee
where $\kappa_0$ is the integration constant. In the weak lensing regime,
Eqs.(9) provides a  mapping of the total projected mass, using the
distortion of the background objects.

\begin{table}
\hskip 2.0 truecm 
\psfig{figure=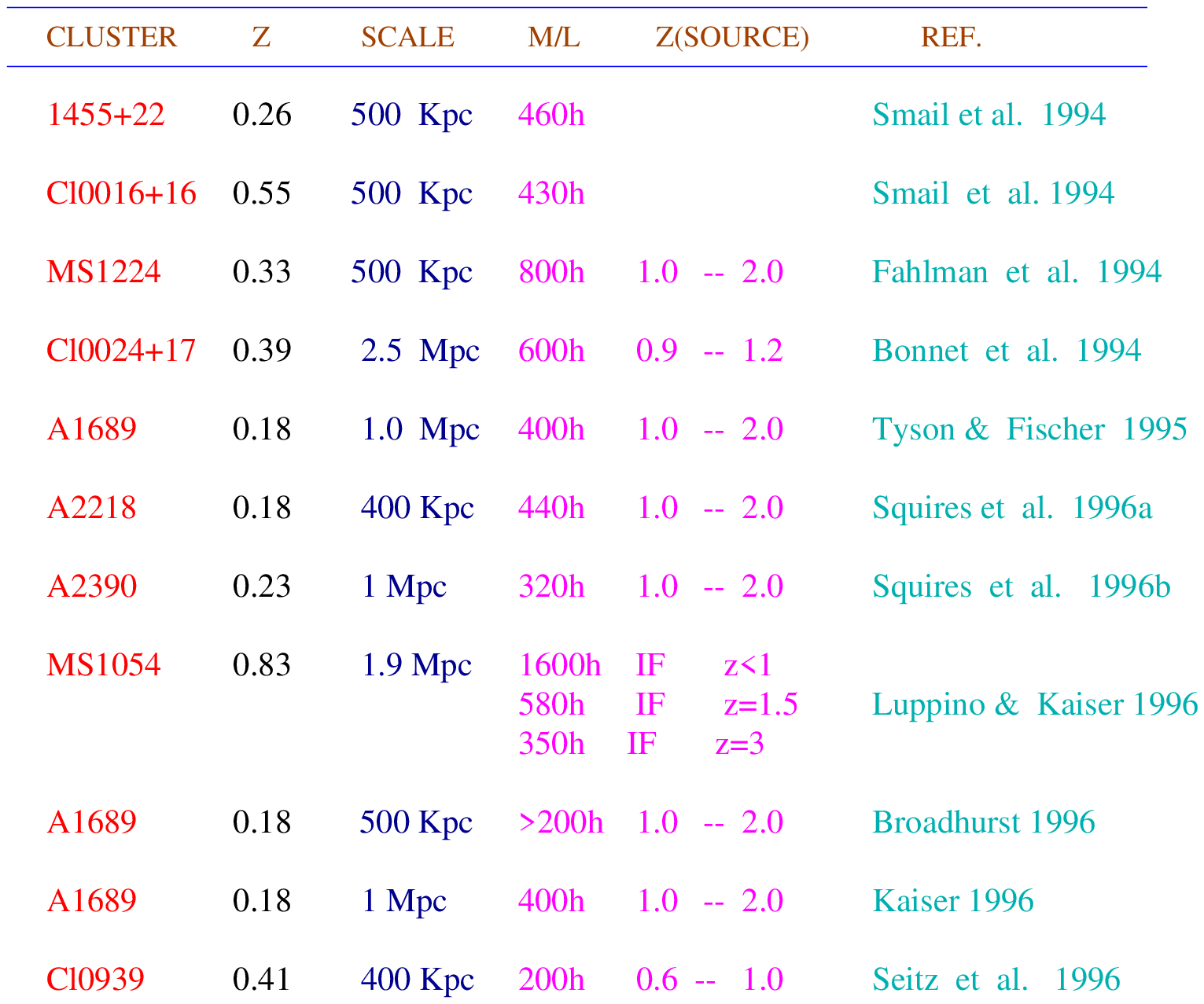,width=14. cm}
\caption{Summary mass obtained from weak lensing inversion in the
literature. The averaged mass-to-light ratio is higher than the one
inferred from giant arc (see table 1) which is interpreted as a real 
increase with distance from the cluster center. Note the strong
uncertainty in the case of MS1054 which is due to the strong dependence
of mass with the redshift of sources as the redshift of the lens
increases. The scales are expressed in $h_{100}^{-1}$ Kpc.}
\end{table}

Several improvements of the basic theory of lensing inversion have been 
discussed in details by Seitz \& Schneider (1995, 1996), Schneider (1995) and 
Kaiser (1996). Note that 
the objects are also magnified by a factor $\mu$ which, in case of the
weak lensing regime, has the form,
\be
\mu=1+2\kappa\ .
\ee
This gives potentially
another independent way to measure the projected mass density
$\Sigma$ of a lens using the magnification instead of the distortion.

Table 2 gives a summary of the clusters for which lensing 
inversion have been attempted. When compare with mass distribution 
inferred from strong lensing, there is a clear trend towards higher $M/L$ 
when the scale increases.  Bonnet et al. (1994) found $M/L$ close to 
600 at 2.5 \hMpc 
from the cluster center. Note also the remarkable results from 
Kaiser \& Luppino on a cluster at redshift 0.83, for which the
estimated mass strongly depends on the redshifts of the sources. 
Indeed in general the estimation of the cluster mass using Eq. (9) requires the
knowledge of $\Sigma_{\rm crit.}$, which depends on the 
usually poorly known redshifts of the
sources. Though this is not a critical issue for
nearby clusters ($z_l<0.2$), because then ${D_{os}/ D_{ls}}\simeq 1$, it could
lead to large mass uncertainties for more distant clusters it is the case
for MS1054.

Actually, even if the redshift of the sources were known, it would 
still not be possible to get 
the absolute value of the mass distribution, because possible
mass planes of constant density intercepting the line of sight 
do not change the shear map.
Mathematically, this corresponds to the
unknown integration constant $\kappa_0$ in Eq.(10).
This degeneracy may be broken if one    
measures the magnification $\mu$ which depends on the mass quantity inside
the light beam (Eq.(3)). While the
shear measurement does not require any information in the source plane, the
magnification measurement needs the observation of a reference (unlensed)
field to calibrate the magnification. Broadhurst et al. (1995) proposed to
compare the number count
$N(m,z)$ and/or $N(m)$ in a lensed and an unlensed field to measure
$\mu$. Depending on
the value of the slope $S$ of the number count in the reference field, we
observe a bias (more objects) or an anti-bias (less objects) in the
lensed field. The particular value $S=0.4$ corresponds to the case where
the magnification of faint objects is exactly compensated by the
dilution of the number count (Eq.(18)). This method was applied on
the cluster A1689 (Broadhurst, 1995), but the signal to noise of the
detection remains 5 times lower than with the distortion method for a given
number of galaxies. The magnification may also be determined
by the changes of the image sizes at fixed surface brightness
(Bartelmann \& Narayan 1995).

The difficulty with these methods is that they
required to measure the shape, size and magnitude of very faint
objects up to B=28
which depends on the detection threshold, the
convolution mask and the local statistical properties of the noise.
These remarks led us to propose a new method to analyze
the weak lensing effects, based on the auto-correlation function of 
the pixels in
CCD images, which avoids shape parameter measurements of individual
galaxies (Van Waerbeke et al. 1996a). It is described in the next
subsection.

\subsection{The Auto-correlation method}
The CCD image is viewed as a density field rather than an image containing
delimited objects. The surface brightness, $I(\vec \theta)$, 
in the image plane in the direction
$\vec \theta$ is related to the surface brightness in the source plane
$I^{(s)}$ by the relation,
\be
I(\vec \theta)=I^{(s)}({\cal A}\vec \theta),
\ee
which can be straightforwardly extended
to the auto-correlation function (ACF) (e.g. the
two-point autocorrelation function of the light distribution in a given area),
\be
\xi(\vec \theta)=\xi^{(s)}({\cal A}\vec \theta)\ .
\ee
This equation is more meaningful when it is written in the
weak lensing regime,
\be
\xi(\vec \theta)=\xi^{(s)}(\theta)-\theta \ \partial_{\theta} \xi^{(s)}(\theta)
[1-{\cal A}]
\ee
since the local ACF, $\xi(\vec \theta)$, now writes as 
the sum of an isotropic unlensed term,
$\xi^{(s)}(\theta)$, an isotropic lens 
term which depends on $\kappa$, and an anisotropic term which depends on
$\gamma_i$.

Let us now explore the gravitational lensing information that can be extracted
from the shape matrix $\cal M$ of the ACF,
\be
{\cal M}_{ij}={\int \d^2\theta\ \xi (\vec \theta)\ \theta_i\ \theta_j\over \int
\d^2\theta\ \xi (\vec \theta)}\ .
\ee
The shape matrix in the image plane is simply related to the shape
matrix in the source plane ${\cal M}^{(s)}$ by ${\cal M}_{ij}={\cal
A}_{ik}^{-1} {\cal A}_{jl}^{-1} {\cal M}^{(s)}_{kl}$. If the galaxies
are isotropically distributed in the source plane, 
$\xi^{(s)}$ is isotropic, and in that case ${\cal
M}^{(s)}_{ij}=M\delta_{ij}$, where $\delta_{ij}$ is the identity matrix.
Using the expression of the amplification matrix $\cal A$ we get the
general form for $\cal M$,
\be
{\cal M}={M(a+|g|^2)\over (1-\kappa)^2(1-|g|^2)} \pmatrix{
1+\delta_1 & \delta_2 \cr \delta_2 & 1-\delta_1 \cr }\ .
\ee
The observable quantities (distortion $\delta_i$ and magnification
$\mu$) are given in terms of the components of the shape matrix,
\be
\delta_1={{\cal M}_{11}-{\cal M}_{22}\over {\rm tr}({\cal M})} \ ; \ \ \
\delta_2={2{\cal M}_{12}\over {\rm tr}({\cal M})} \ ; \ \ \
\mu=\sqrt{{\rm det}({\cal M})\over M},
\ee
where ${\rm tr}({\cal M})$ is the trace of $\cal M$ and
${\rm det}({\cal M})$ is the determinant of $\cal M$. 
As for sheared galaxies, we see that the
distortion is available from a direct measurement in the image plane while
the magnification measurement requires to know the value of $M$ which
is related to the light distribution in the source plane, or in an
unlensed reference plane. The ACF provides a new and independent 
way to measure
$\delta_i$ and $\mu$ which does not require shape, size or photometry
of individual galaxies. Furthermore, the signal to noise ratio is 
proportional to the number density of background galaxies, $N$, instead 
as $\sqrt{N}$ for the standard method (see figure 2).


A description of its practical implementation and first results are given 
in Van Waerbeke et al. (1996a) and  Van Waerbeke \& Mellier (1996b). Clearly, 
the ACF is the most powerful technique for measuring  the orientation of 
very weak shear as the one we
expect from large-scale structures.

\section{The matter distribution on very large scale}

The direct observation of the mass distribution on scale larger than 
10 \hMpc (or $\approx $ 1 square degree) is one of the great hopes of 
the weak gravitational lensing approach.  Two observational directions are now 
being investigated. In the first one we search for the shear of dark
condensation of mass that can be responsible for the magnification bias of
luminous quasars and radio-sources. It probes the lumpiness of matter
distribution within large-scale structures.  In the second one, we analyze
the statistical properties of weakly lensed background galaxies on
degree scales in order to
obtain constraints on the cosmological parameters and  
on the projected power spectrum.

\subsection{Shear around radio-sources and the lumpiness of matter}

Fugmann (1990) and  Bartelmann \& Schneider (1993a,b) have 
demonstrated that  there is a strong correlation  between the presence of
 galaxies or clusters and bright quasars. They interpreted this as a
 magnification  bias  induced the galaxy over-densities along the line of 
sight.  However, the measured correlation is too strong to be only due to 
individual galaxies, so  Bartelmann \& Schneider suggested that the 
magnification 
bias originates from groups or even rich cluster of galaxies. 
If this suggestion is correct, the deflector responsible for the 
magnification bias should also induce gravitational shear onto background 
sources.

\begin{figure}
\hskip 3truecm
\psfig{figure=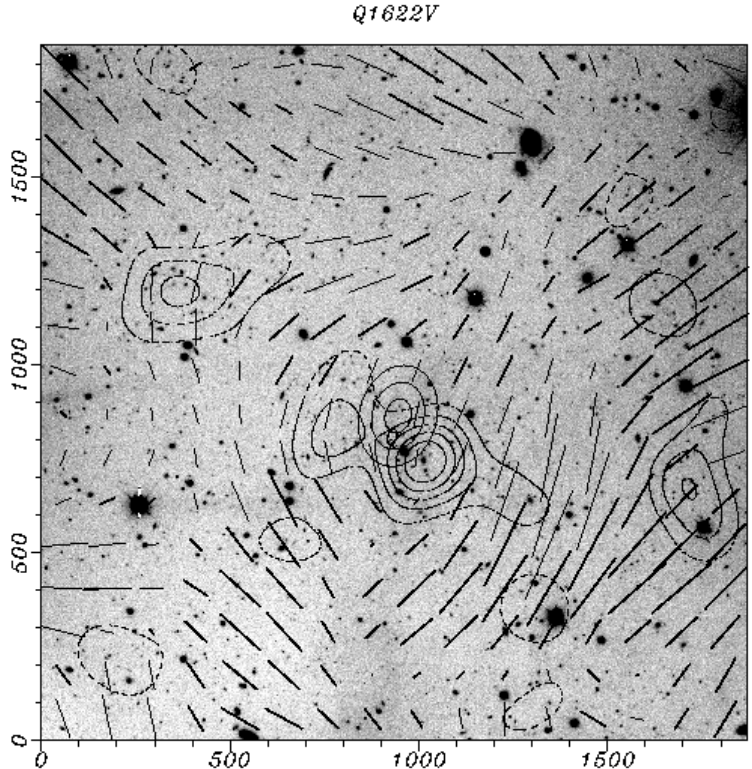,width=10. cm}
\caption{Shear map around the bright quasar Q1622. The image obtained at
CFHT clearly shows a coherent shear. The ellipses shows the center of
the shear pattern which is very close to the quasar (dark dot). The 
 others solid lines are galaxy number isodensity contours. Clearly the quasars,
the shear pattern and a galaxy concentration are almost positioned at
the same place which reinforces the hypothesis of a magnification effect 
on the quasar.}
\end{figure}

Bonnet et al. (1993), Mellier et al. (1994) and 
Van Waerbeke et al. (1996a) found a strong gravitational shear  
and a galaxy excess around the quasar pair Q2345+007, the shear 
pattern being associated with a group or a small cluster.  
In the same way Fort et al. (1996a) have measured the shear around a few 
over-bright QSO-s that could be magnified by nearby lensing mass (figure
3).  It is the first tentative to detect
gravitational structures from a mass density criteria rather than 
luminosity excess. The field of view around each QSO is small 
and up to now, there is not enough data to draw a synthetic view of the
global matter distribution. However, both weak lensing detection around
bright radio-sources and the correlations found by Bartelmann \&
Schneider seem to favor a model where clumps of dark matter are more
numerous that expected and concentrated
on groups and clusters of galaxies.  

\subsection{Statistical analysis of shear on very large scale}

In the statistical studies of the shear at very large scale 
the lenses are not individually identified, but viewed as
a random population affecting the shape of the galaxies with
an efficiency depending on their distances.
Indeed, in this approach, we 
consider the statistical
properties of the shear measured on backgroung objects
for randomly chosen line-of-sights. The measured shears 
are actually filtered at a given angular scale so that a signal of
cosmological interest can be extracted.
For a filtering scale of about one degree, the structures
responsible of the gravitational shear being at a redshift
of about 0.4 are expected to be
on scales above 10 \hMpc, that is in a regime where their properties 
can be easily predicted with the linear or perturbation theory.
Within this line of thought, 
Blandford et al. (1991), Miralda-Escud\'e (1991) and Kaiser 
(1992) argued that the projected power spectrum 
should be measurable with such a method
provided shape parameters are averaged on the degree scale, as 
it is illustrated on figure 4.

\begin{figure}
\hskip 2.0 truecm
\psfig{figure=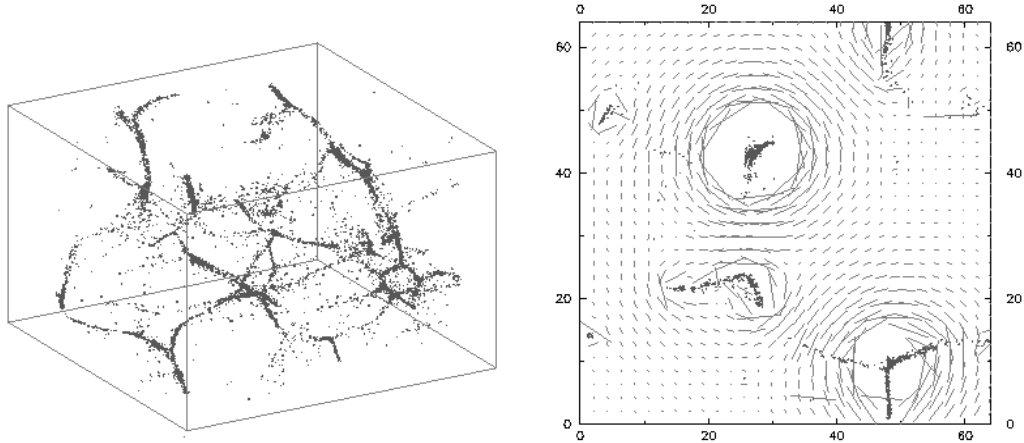,width=14. cm}
\caption{
Simulation of shear expected from large scale structures. The left 
panel shows a 256$^3$ Mpc$^3$ box with dark dots indicating the
location of matter. The long filaments are large scale structures 
which originated from a uniform distribution and an initial power
spectrum $P(k)=k^{-1}$ under the adhesion approximation.  The right
panel shows a slice which represents the projected mass distribution 
as it would be observed by an observer. The thin straight lines are the
local orientation and intensity of the shear. It illustrates what we
expect from the future observations of weak lensing with wide field 
CCD.
}
\end{figure}

For such a scale, however, the shear is expected to be as low as 1\%.
Therefore 
its detection requires high image quality to avoid uncontrolled errors, 
and large angular coverage so that it is possible to separate the
gravitational shear from the intrinsic galactic ellipticities
by averaging over thousands of galaxies. So far such a detection has
not been completed due to these severe constraints, but
there are now projects to build very large CCD mosaic camera which could be
capable of doing large deep imaging surveys. The MEGACAM project conducted by 
the French institutions, CEA (CE in Saclay) and INSU, 
the Canadian CNRC and the CFHT Corporation consists in building a 
16K$\times$16K camera at the prime focus of CFHT. This camera will
provide a total field of view of 1$^o\times$1$^o$, with image quality 
 lower than 0.5" on the whole field.  Using a Pertubation Theory
approach, Bernardeau et al. (1996) have 
analyzed the statistical properties of the gravitational shear averaged
on MEGACAM scales. They have shown that  with the observation of 
about 25 such fields,
one could recover the projected power spectrum and $\Omega$
independently by using the variance and the skewness (third moment)
of the one-point probability distribution function
of the local convergence in the sample
\footnote{detailed calculations have shown however that there is a slight 
degeneracy with the cosmological constant.}. 
Basically, these moments write,
\be
\mg\kappa^2_{\theta}\md \propto \ \ P(k)\ \Omega^{1.5} \ \ z_s^{1.5}
\ee
and
\be
{\mg\kappa^3_{\theta}\md \over \mg\kappa^2_{\theta}\md^2} \propto 
\ \ \Omega^{-0.8} \ \ z_s^{-1.35}
\ee
where $\theta =30'$ is the scale where the convergence $\kappa$ is
averaged, $P(k)$ is the projected power spectrum of the dark matter
and $z_s$ is the
averaged redshift of sources. Note that the skewness, expressed
as the ratio of the third moment by the square of the second, 
does not depend on
$P(k)$ and provides a direct information on $\Omega$.  

These important results show that a very large-scale survey of
gravitational shear can provide important cosmological results 
which could be compared with those coming from large-scale flows and 
observations of the cosmic microwave background anisotropy spectrum
that are expected to culminate with the 
COBRAS/SAMBA satellite mission. However, there are two shortcomings that must
be handled carefully: first, it requires detection of very weak shear
from shape parameters of galaxies. Thus, the image quality of the survey is 
a crucial issue. The ACF method proposed by Van Waerbeke et al. (1996a)
should provide accurate shape information with a high signal to noise
ratio. It is probably
the best available method for measuring shear on very large
scales.  The second point is the strong dependence of the variance and the
skewness with the redshift of lensed sources. Since there are no hopes to
obtain redshifts of these galaxies even from spectroscopy with the VLT-s,
we face on a difficult and crucial problem. It may be overcome 
if multicolor photometric data can provide
accurate redshifts for very faint galaxies. Another possible solution
could be 
the analysis of the radial magnification bias of faint distant galaxies
around rich clusters which, indeed, probes the   
redshift distribution of the sources (Fort et al. 1996b and see section 6).

\section{Measuring the cosmological constant}

One remarkable property of gravitational lensing effect is the local change 
of the galaxy number density. The observable number density of sources
results of the competition between deflection 
effect which tends to 
enlarge the projected area and the magnification which increases 
the number of faint sources. The expected galaxy number density is,
\be
N(r) = N_0\,\mu(r)^{2.5 \alpha -1},
\ee
where $\mu$ is the magnification at the position $r$ and 
$\alpha= {\rm d}\ \log(N)  / {\rm d} m$ 
is the slope of the galaxy counts. When $\alpha$ is larger 
than 0.4, the galaxy number density increases. At faint limiting magnitude, 
$\alpha$ becomes significantly smaller than 0.4 and a clear galaxy depletion 
can be detected (Broadhurst 1995, Fort et al. 1996b). 

\begin{figure}
\hskip .5 truecm
\psfig{figure=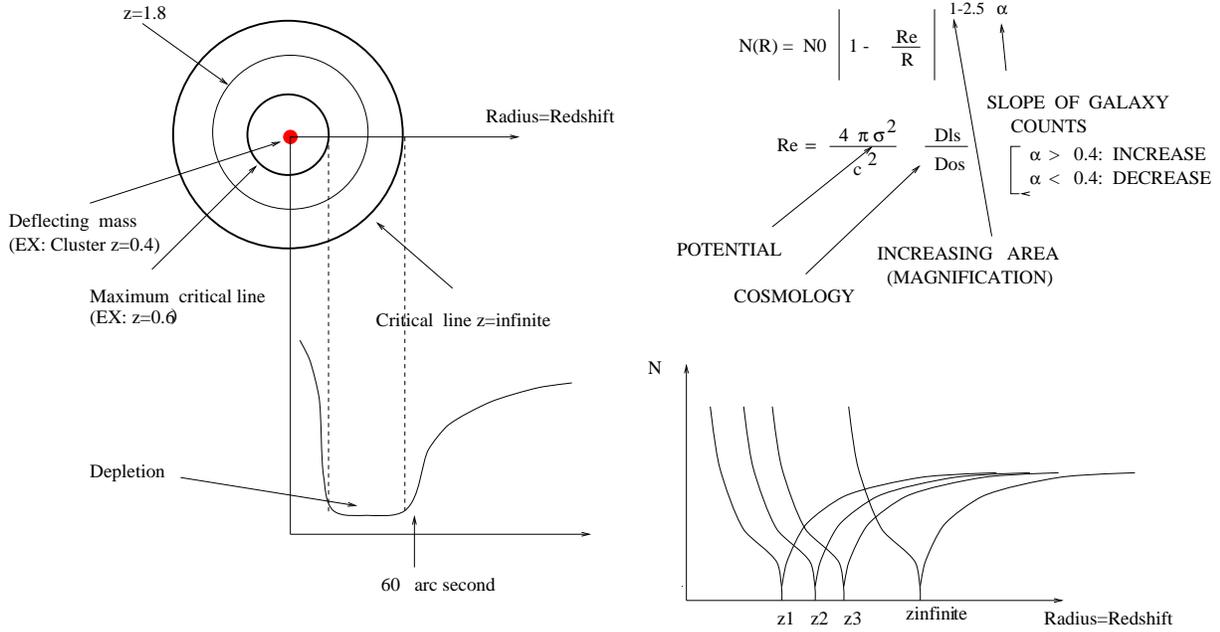,width=16. cm}
\caption{Principle of measurement of redshifts from depletion. The 
right panel shows the depletions curves expected by a singular
isothermal sphere. If the lens is perfectly known 
the minimum of each depletion curve (right panel) depends only 
of the redshift of the source
and the radial position of the critical line is actually
equivalent to a redshift. 
In a realistic case, the redshift 
distribution is broad and the left  panel shows the depletion as it would be
observed: instead of
the single peaked depletion we expect a more pronounced minimum  between
two radii (= two redshifts ) whose angular positions strongly depend on the
cosmological constant for high redshift sources. Thus, if the mass
distribution of the lens is well known as in the rich lensing cluster 
Cl0024+1654, $\lambda$ can be inferred from the shape of the depletion curve.  
 }
\end{figure}

\begin{figure}
\psfig{figure=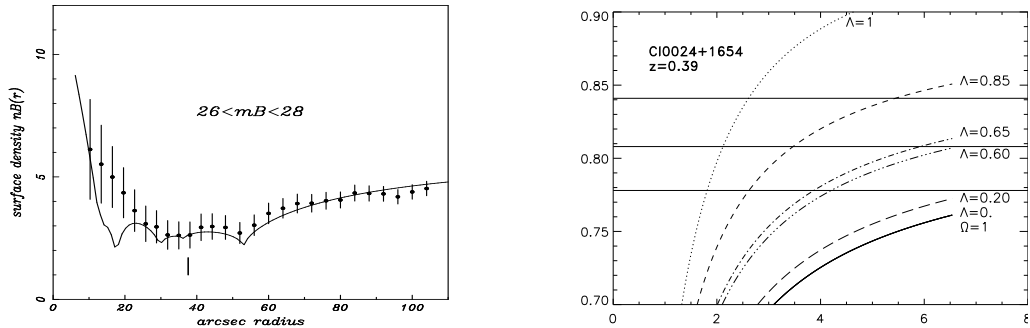,width=14. cm}
\caption{Measuring the cosmological constant from the depletion observed
in Cl0024. The left panel shows the depletion curve observed in
Cl0024+1654. From the redshift of the first minimum and the second
minimum, one can constrain the cosmological constant in order to
position the second minimum at its right angular position. Whatever the
redshift of the most distant sources visible on the images, we see that
the angular position where the depletion curve raises again imposes that
$\lambda>0.65$}
\end{figure}

The shape of the depletion curves depends on the magnification $\mu$ 
which is a complicated function of the lensing potential, the redshift
distribution of sources and the cosmology. For a singular isothermal 
sphere, the amplification at radius $r$ writes,
\be
\mu(r) = {4 \pi \sigma^2 \over c^2}\,{D_{ls} \over D_{os}}\,{r \over r-1},
\ee
and the depletion curve for a single redshift of sources looks like
those shown in figure 5.  For a redshift distribution, the depletion curve 
shows 
a plateau between two extremum points corresponding to the lowest 
and highest redshift of the sources. Once the redshift of one source 
is known, the radial position of the highest redshift strongly depends
on the cosmological constant. 

Fort et al. (1996b) tentatively measured depletions curves 
in Cl0024+1654 (see figure 6) and A370 of faint galaxies close to the noise 
level.
With the hypothesis of a single lens along the line of sight they 
found that the angular position of the highest-redshift sources is
very high and imposes that the most distant galaxies visible in the 
field have redshifts larger than $2$, while the width of the depleted areas 
extend as far as 60 arcseconds which is incompatible with a low-$\lambda$ 
universe. In fact, the observations provide a lower limit  $\lambda >0.6$ 
(figure 6). 

\section{Conclusion}

In the last ten years, 
the gravitational lensing effects turned out to be among the most 
promising tools for cosmology. It is indeed a direct probe
of the large-scale cosmic mass
distribution and some  observable quantities revealed to be
extremely sensitive  to
the cosmological parameters. We summarize in table 3 
what we learned about $\Omega$ from the mass distribution in rich clusters 
of galaxies and what we expect in the near future.  It shows that we
can hope for strong and reliable  constraints on $\Omega$ and $\lambda$
from the developing observational tools:
we are now aiming at a determination of the
cosmological parameters within 10\% accuracy. 

\begin{table}
\hskip 3.0 truecm
\psfig{figure=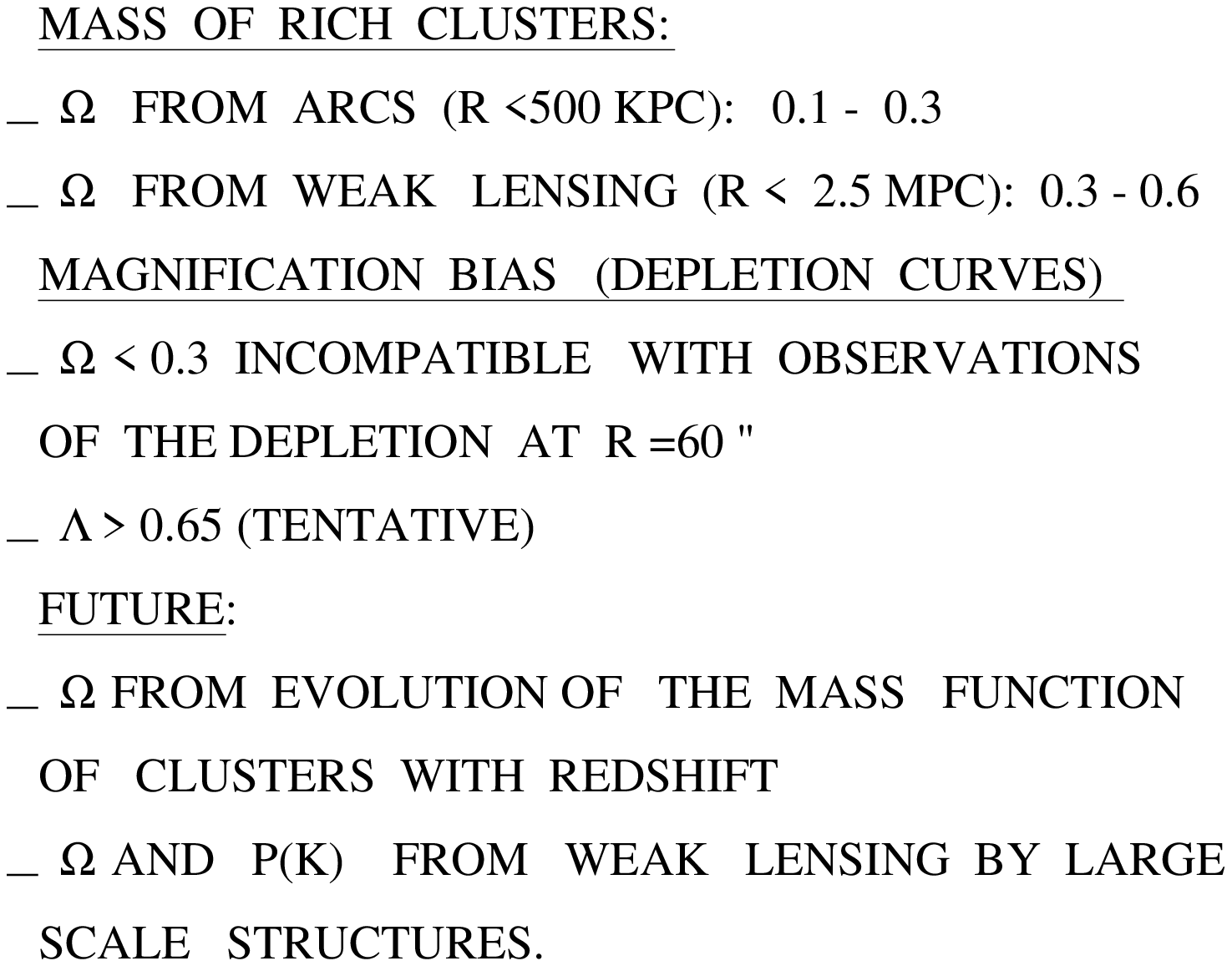,width=10. cm}
\caption{A summary of the values of the cosmological parameters as 
they are inferred from gravitational lensing. They are still some 
uncertainties and many hopes for the future. But a real trend toward
$\Omega >0.3$ seems well established.}
\end{table}



\acknowledgements
We thanks P. Schneider, for discussions and
enthusiastic support.
F. Bernardeau is grateful to IAP, where most of this work has been completed,
for its hospitality.

\section{References}

{\parindent=0pt
\parskip=3 pt

Bartelmann, M., Schneider, P. (1993a) A\&A 259, 413.

Bartelmann, M., Schneider, P. (1993b) A\&A 271, 421.

Bartelmann, M. (1995) A\&A 299, 661.

Bartelmann, M., Narayan, R. (1995) ApJ 451, 60.

Bernardeau, F., Van Waerbeke, L., Mellier, Y. (1996) in preparation

Blandford, R.D., Saust, A.B., Brainerd, T.G., Villumsen, J.V. (1991),
MNRAS, 251, 600

Bonnet, H., Fort, B., Kneib, J-P., Mellier, Y., Soucail, G. (1993) A\&A
280, L7

Bonnet, H., Mellier, Y., Fort, B. (1994) ApJ 427, L83.

Bonnet, H., Mellier, Y. (1995) A\&A 303, 331.

Broadhurst, T. (1995) SISSA preprint astro-ph/9511150.

Broadhurst, T., Taylor, A.N., Peacock, J. (1995) ApJ 438, 49.

Fahlman, G., Kaiser, N., Squires, G., Woods, D. (1994) ApJ 437. 56.

Fort, B., Prieur, J.-L., Mathez, G., Mellier, Y., Soucail, G. (1988)
A\&A 200, L17.

Fort, B., Mellier, Y., (1994) A\&A Review 5, 239, 292.

Fort, B., Mellier, Y., Dantel-Fort, M., Bonnet, H., Kneib, J.-P. (1996a)
A\&A 310, 705.

Fort, B., Mellier, Y., Dantel-Fort (1996) SISSA preprint
astro-ph/9606039.

Fugmann, W. (1990) A\&A 240, 11.

Kaiser, N. (1992) ApJ 388, 272.

Kaiser, N. (1996) SISSA preprint astro-ph/9509019.

Kaiser, N., Squires, G. (1993) ApJ 404, 441

Kneib, J.-P., Mellier, Y., Fort, B., Mathez, G. (1993) A\&A 273, 367.

Kneib, J.-P., Mellier, Y., Pell\'o, R., Miralda-Escud\'e, J., Le
Borgne, J.-F., Boehringer, H., Picat, J.-P. (1995) A\&A 299, 168.

Kassiola, A., Kovner, I., Fort, B. (1992) ApJ 400, 41.

Luppino, G., Kaiser, N. (1996) SISSA preprint astro-ph/9601194.

Lynds, R., Petrossian, V. (1986) BAAS 18, 1014.

Mellier, Y., Fort, B., Kneib, J.-P. (1993) ApJ 407, 33.

Mellier, Y., Dantel-Fort, M., Fort, B., Bonnet, H. (1994) A\&A 289,
L15.

Miralda-Escud\'e, J., (1991) ApJ, 370, 1.

Narayan, R., Bartelmann, M. (1996) SISSA preprint astro-ph/9606001

Pierre, M., Le Borgne, J.-F., Soucail, G., Kneib, J.-P. (1996) A\&A 311,
413.

Schneider, P., Ehlers, J., Falco, E. E., (1992), {\it Gravitational
Lenses}, Springer.

Seitz, C., Kneib, J.P., Schneider, P., Seitz, S., (1996) in press. 

Seitz, C., Schneider, P., (1995) A\&A 302, 9

Seitz, S., Schneider, P., (1996) A\&A 305, 388 

Smail, I., Ellis, R.S., Fitchett, M. (1994) MNRAS, 270, 245.

Schneider, P. (1995), A\&A, 302, 639

Soucail, G., Fort, B., Mellier, Y., Picat, J.-P. (1987) A\&A 172, L14. 

Squires, G., Kaiser, N., Babul, A., Fahlmann, G., Woods, D., Neumann,
D.M., B\"ohringer, H. (1996a) ApJ 461, 572.

Squires, G., Kaiser, N., Falhman, G., Babul, A., Woods, D. (1996b) SISSA
preprint astro-ph/9602105.

Tyson, J.A., Fisher, P. (1995) ApJL, 349, L1.

Van Waerbeke, L., Mellier, Y., Schneider, P., Fort, B., Mathez, G.  (1996a)
A\&A in press. SISSA preprint astro-ph/9604137.

Van Waerbeke, L., Mellier, Y. (1996b). Proceedings of the XXXIst
Rencontres de Moriond, Les Arcs, France 1996. SISSA preprint
astro-ph/9606100.

Wallington, S., Kochanek, C. S., Koo, D. C. (1995) ApJ 441, 58.

Walsh, D., Carswell, R.F., Weymann, R.J., (1979), Nature, 279, 381
}

\newpage
\heading{PROGR\`ES R\'ECENTS EN LENTILLE GRAVITATIONNELLE}

\centerline{Y. Mellier$^{(1,4)}$, L. Van Waerbeke$^{(2)}$, F. Bernardeau$^{(3)}$,
B. Fort$^{(4)}$}
{\it
\centerline{$^{(1)}$Institut d'Astrophysique de Paris,}
\centerline{98$^{bis}$ Boulevard Arago,}
\centerline{75014 Paris, France.}
\centerline{$^{(2)}$ Observatoire Midi Pyr\'en\'ees,}
\centerline{14 Av. Edouard Belin,}
\centerline{31400 Toulouse, France.}
\centerline{$^{(3)}$ Service Physique Th\'eorique,}
\centerline{CE de Saclay,}
\centerline{91191 Gif-sur-Yvette  Cedex, France.}
\centerline{$^{(4)}$Observatoire de Paris (DEMIRM),}
\centerline{61 Av. de l'Observatoire,}
\centerline{75014 Paris, France.}
}

\begin{abstract}{\baselineskip 0.4cm 
L'effet de lentille gravitationelle est aujourd'hui consid\'er\'e comme
un des outils les plus prometteurs de la cosmologie.  Il sonde
directement la mati\`ere distribu\'ee dans les grandes structures 
et peut aussi fournir des informations importantes sur les param\`etres
cosmologiques, $\Omega$ et $\Lambda$.  Dans cet article de revue, nous
r\'esumons les progr\'es observationels et th\'eoriques r\'ecents les plus 
marquants  obtenus dans ce domaine au cours des cinq derni\`eres
ann\'ees.
}
\end{abstract}
\end{document}